\documentclass[twocolumn,showpacs,amsmath,amssymb]{revtex4}
\usepackage[utf8]{inputenc}
\usepackage{color}
\usepackage{bm}
\usepackage{amsmath}
\usepackage{graphicx}
\usepackage{esint}

\makeatletter

\@ifundefined{textcolor}{}
{%
 \definecolor{BLACK}{gray}{0}
 \definecolor{WHITE}{gray}{1}
 \definecolor{RED}{rgb}{1,0,0}
 \definecolor{GREEN}{rgb}{0,1,0}
 \definecolor{BLUE}{rgb}{0,0,1}
 \definecolor{CYAN}{cmyk}{1,0,0,0}
 \definecolor{MAGENTA}{cmyk}{0,1,0,0}
 \definecolor{YELLOW}{cmyk}{0,0,1,0}
}

\makeatother

\begin{document}
\title{Designing Autonomous Maxwell's Demon via Stochastic Resetting}
\author{Ruicheng Bao}
\thanks{These authors contributed equally to this work.}
\author{Zhiyu Cao}
\thanks{These authors contributed equally to this work.}
\author{Jiming Zheng}
\thanks{These authors contributed equally to this work.}
\author{Zhonghuai Hou}
\thanks{E-mail: hzhlj@ustc.edu.cn}
\affiliation{Department of Chemical Physics \& Hefei National Laboratory for Physical
Sciences at Microscales, iChEM, University of Science and Technology
of China, Hefei, Anhui 230026, China}
\date{\today}
\begin{abstract}
Autonomous Maxwell's demon is a new type of information engine proposed
by Mandal and Jarzynski {[}Proc. Natl. Acad. Sci. U.S.A. 109, 11641
(2012){]}, which can produce work by exploiting an information tape.
Here, we show that a stochastic resetting mechanism can be used to
improve the performance of autonomous Maxwell's demons notably. Generally,
the performance is composed of two important features, the time cost
for an autonomous demon to reach its functional state and its efficacious
working region in its functional state. Here, we provide a set of
design principles for the system, which are capable of improving the
two important features. On the one hand, one can drive any autonomous
demon system to its functional periodic steady state at a fastest
pace for any initial distribution through resetting the demon for
a predetermined critical time and closing the reset after that. On
the other hand, the system can reach a new functional state when the
resetting is always on, in which case the efficacious region of the
demon could be extended significantly. Moreover, a dual-function region
in a new phase diagram of the demon with resetting has been found.
Remarkably, in this dual-function region the demon with resetting
can realize anomalous output of work and erasure of information on
the tape simultaneously, violating the second law of thermodynamics
apparently. To this question, we derive a new modified Clausius inequality
to restore the second law by taking the cost of resetting into account. 
\end{abstract}
\maketitle

\section{Introduction}

In 1867 \cite{Maxwell1871}, James C. Maxwell conceived an intelligent
creature to challenge the second law of thermodynamics, which usually
refers to as ``Maxwell's demon'' nowadays. The demon is capable
of doing work only through harvesting information, which seems to
violate the second law. This puzzle was finally clarified by Landauer
and Bennet \cite{landauer1961irreversibility,bennett1982thermodynamics}.
What the crucial point is that information needs to be stored in a
physical memory, so that during each thermodynamic cycle an amount
of energy should be cost to reset the memory, which promises the validity
of second law. Maxwell's thought experiment led to a series of interdisciplinary
studies focusing on the interrelation between information theory and
small-system thermodynamics. Recent years have witnessed great progress
in this field named as information thermodynamics, including fruitful
experimental works \cite{2017Maxwell} and theoretical studies \cite{2015Thermodynamics}. 

Over the past 150 years, Maxwell's-demon-like models have been subjected
to extensive theoretical study \cite{szilard1929entropieverminderung,feynman2011feynman,mandal2012work,2015Thermodynamics}.
Most of these models can be roughly divided into two classifications,
one is measurement-feedback-controlled demon and another is autonomous
demon. On the one hand, T. Sagawa and M. Ueda have developed the theoretical
formalism of the first class of measurement-feedback demon \cite{2012Nonequilibrium}.
On the other hand, Mandal and Jarzynski have constructed two analytically
solvable autonomous demon models \cite{mandal2012work,mandal2013maxwell},
both consist of a memory-tape and a demon. These memory-tape autonomous
demons can realize processes that are forbidden by the second law
of thermodynamics by exploiting information. There are many subsequent
works about these two kinds of Maxwell's demon models \cite{AMD_14PRE,AMD_15experiment,AMD_20PRR_Crutchfield,AMD_21PRE,AMD_22arxiv_Jarzynski,AMD_22PRE_finite,MD_14PRE_continuous,MD_21PRL_Gambling,MD_22arxiv_Esposito}.
For instance, both of these two demon models are further generalized
to quantum system \cite{22NJP_QMD,13PRE_QMD,22PRE_QMD}. 

Stochastic resetting, a rather common driving mechanism that stops
a dynamical process randomly and starts it anew, has recently been
of great interest in the field of statistical physics \cite{11PRL_reset,14PRL_reset,16PRL_reset,17PRL_Rev_reset,18PRL_reset_Sergey,18PRL_reset_Sokolov,19PRL_reset_Arnab,21PRR_reset_Rev,22PRL_resetBridge,20PRL_FPresetting}.
It has been shown that restarting a dynamical process stochastically
may accelerate the process on average, which is counterintuitive.
For instance, Evans and Majumdar \cite{11PRL_reset} studied stochastic
resetting theoretically for the first time,~ claiming that the mean
time for a freely diffusive Brownian particle to find a fixed target
turns to be finite under constant rate Poisson resetting, while the
mean time diverges without resetting. From then on, the resetting
mechanism has been demonstrated to be advantageous to plenty of different
stochastic processes such as animal foraging, RNA polymerases backtrack
recovery \cite{16PRE_reset} and relaxation processes \cite{Busiello_2021}.
Furthermore, the effect of stochastic resetting on thermodynamics
is also of interest \cite{20PRL_workreset,20PRR_EPreset,20Speedlimit_reset,21PRR_TURreset,2016reset_thermal,IFT_reset}.
For more discussions of stochastic resetting see two recent reviews
\cite{2019Stochastic,2022reset}. Generally, this stochastic resetting
mechanism could be used to optimize controlling protocols in small
systems. 

There are two important features of autonomous demons, which are
main issues pertinent to the evaluation of their performances. The
first feature is the relaxation time scale for the demon to reach
its functional periodic steady state \cite{22arXiv_ZyCao}. It would
be more accessible to initially prepare the demon in an equilibrium
state which is generally not the periodic steady state, so there is
some time cost for the demon to enter its working state and start
transforming information resource into work. Certainly, the less time
cost due to demon's initial deviation from the functional state is
more preferable. Another crucial feature is the demon's efficacious
working regions, including the region where positive work can be produced
by exploiting information resources (information engine region) and
the region in which the demon serve as a eraser that can replenish
information resources (information eraser region, where the information
entropy of the memory tape decreases on average). One would expect
to extend those two working regions of the demon for better performance.

Here in the present work, we utilize the discrete-time stochastic
resetting mechanism to provide designing principles of the autonomous
Maxwell's demon, improving demon's performance in the two aspects
of relaxation time-scale and efficacious working regions. To regulate
those two main features of the autonomous demon, we introduce two
kinds of resetting strategies leading to the reduction of time cost
and the extension of the useful working regions respectively. The
first type of strategy is to reset the demon to a given state with
constant probability in discrete time (at the beginning of each cycle)
for a fixed amount of time $t_{c}$, then turn off the reset and let
the demon evolves according to the original dynamics. Strikingly,
this can drive the demon to reach its functional periodic steady state
at the fastest pace whatever its initial distribution is. This significant
acceleration strategy is inspired by the so-called ``strong'' Mpemba
effect, which recently received a lot of attention \cite{gal2020precooling,klich2019mpemba,kumar2020exponentially,lu2017nonequilibrium,santos2020mpemba,baity2019mpemba}.
The second type of strategy is to keep resetting the demon to a given
state, then the demon system will finally arrive at a new periodic
steady state depending on the resetting rate. The new periodic steady
state with resetting shows properties different to the original autonomous
demon. In the phase diagram of this new autonomous demon with resetting,
the information engine region is extended and the information eraser
region is almost unchanged meanwhile. Remarkably, we find that there
is an overlapping part of the information engine region and the information
eraser region in the new phase diagram. We call this region the 'dual-function
region', where the second law of thermodynamics is apparently violated.
To restore the second law, we derive a new Clausius inequality containing
an extra term from the effect of stochastic resetting, which restores
the second law of thermodynamics in the presence of resetting.

This article is organized as follows. In section II, we describe the
methods to analyze autonomous Maxwell's demon models introduced by
Mandal and Jarzynski. Moreover, the formalism of discrete time stochastic
resetting is introduced. In section III, we discuss the approach to
induce significantly faster relaxations of the autonomous demon to
functional periodic steady states via stochastic resetting strategy.
New phase diagrams of an autonomous demon with the resetting mechanism
always on is shown in section IV, where the anomalous work region
is extended and the remarkably interesting 'dual-function' region
appears. In section V, we make some discussions and conclude the paper.

\section{Model and framework}

\paragraph*{Setup and analysis of the model}

An autonomous Maxwell's demon consists of a demon with $k$ states
and an infinitely long memory tape (a stream of bits) encoding information
with bit $0$ and $1$. In our setup, the $k$ - state demon is initially
in equilibrium with a thermal reservoir at temperature $T_{\text{in}}$,
then it would be coupled to a memory tape to constitute a $2k$ states
combined system. This memory tape plays the roles of measurement and
feedback as in the conventional Maxwell's demon system. It moves through
the demon at a constant speed to a given direction, with the bit sequence
in the tape being written in advance (e.g. $0$11100...). The bit
sequence is described by a probability distribution $\bm{p}_{\text{in}}^{B}=(p_{0},p_{1})^{\text{T}},$
where $p_{1}$ and $p_{0}$ are the probabilities of the incoming
bit to be in states $1$ and $0$. For later use, we define $\delta\equiv p_{0}-p_{1}$
as the proportional excess of $0$'s among all incoming bits in the
tape. The demon interacts with the incoming bits one by one as they
pass by, i.e., it only interacts with the nearest bit for a fixed
time $\tau$, after that the current bit leaves and a new bit comes
in. During each interaction interval, there are intrinsic transitions
between some pair of states of the demon. Moreover, the bit can transit
between state $0$ and $1$ with the demon transiting between a state
and another state meanwhile, and these type of cooperative transitions
are not allowed to happen in the absence of either demon or bit. These
cooperative transitions could bring about anomalous work production
because the disordered transitions (just like fluctuations) of the
$k$-state demon may be rectified by the incoming bits, which is the
key idea of the autonomous demon. If the outgoing bit stream becomes
more disordered than the incoming bit stream (the information entropy
of the bit increase), then the transition of the demon is rectified
to a given ``direction'' on average. The transition of demon to the
given ``direction'' can produce work at the cost of information
resources, while the opposite ``direction'' is useless. Which ``direction''
of transition can bring about positive work depends on the setup of
the combined system. Throughout this paper, we set the bit transition
from $0$ to $1$ as the right direction which can rectify the demon's
transitions , producing positive output work consequently.

Before proceeding, we list some notations for after use. Importantly,
$t_{N}:=N\tau$ is the beginning time of the $N^{th}$ interaction
interval, with $\tau$ being the interaction time of each interval.
$\bm{p}^{D}(t)$ is a column vector with $k$ entries, which describes
the probability distribution of the demon at time $t$. Similarly,
$\bm{p}^{B}(t)$ is a vector with 2 entries, denoting the state of
the bit at time $t$. $\bm{p}_{\tau}^{D}(t_{N})$ and $\bm{p}_{\tau}^{B}(t_{N})$
are the distributions of the demon and bit at the end of the $N^{th}$
interval, compared to $\bm{p}_{\text{in}}^{D}(t_{N})$ and $\bm{p}_{\text{in}}^{B}(t_{N})$
who are the distributions at the start of the $N^{th}$ interval.
$\boldsymbol{p}(t)$ is the statistical state of the combined system
with $2k$ entries. Finally, $\bm{p}_{\text{in}}^{D,ps}$, $\bm{p}_{\text{\ensuremath{\tau}}}^{D,ps}$and
$\bm{p}_{\tau}^{B,ps}$ are the periodic steady state distributions
at the beginning of intervals of the demon, and at the end of the
intervals of the demon and the bit. The statistical state of the combined
system comprised of the $k$ - state demon and the current bit (with
two states $0$ or $1$) of the tape at the beginning time of the
$N^{th}$ interval is given by the $2k$ dimensional vector 
\[
\boldsymbol{p}_{\text{in}}(t_{N})=\mathcal{M}\bm{p}_{\text{in}}^{D}(t_{N}),\ \ \text{\ensuremath{\mathcal{M}=}}\left(\begin{array}{c}
p_{0}\mathbb{I}\\
p_{1}\mathbb{I}
\end{array}\right),
\]
where $\mathbb{I}$ is a $k\times k$ identity matrix and $\mathcal{M}$
is a $2k\times k$ matrix denoting a mapping from the demon subspace
($k\times1$) to the total combined space ($2k\times1$).What's more,
through defining some projectors, it would be also easy to extract
the distributions in demon subspace and bit subspace from the distributions
in the combined space at any time $t$, as follow: 
\begin{align}
\bm{p}^{D}(t) & =\mathcal{P}^{D}\boldsymbol{p}(t),\nonumber \\
\mathcal{P}^{D} & =(\mathbb{I},\mathbb{I})\\
\bm{p}^{B}(t) & =\mathcal{P}^{B}\boldsymbol{p}(t),\nonumber \\
\mathcal{P}^{B} & \equiv\left(\begin{array}{cccccc}
1 & ... & 1 & 0 & ... & 0\\
0 & ... & 0 & 1 & ... & 1
\end{array}\right)_{2\times2k},
\end{align}
with $\mathcal{P}^{D}$ and $\mathcal{P}^{B}$ denoting the projectors
from the combined space to the demon subspace and to the bit subspace
respectively. The combined system of the demon and the tape evolves
under the master equation 
\begin{equation}
\frac{d}{dt}\boldsymbol{p}(t)=\boldsymbol{R}\boldsymbol{p}(t)\label{eq:evoeqn}
\end{equation}
during an interval from $t=N\tau$ to $t=(N+1)\tau$, where $\boldsymbol{R}$
is a $2k\times2k$ transition matrix whose diagonal elements are $R_{ii}=-\sum_{i\neq j}R_{ji}$,
and off-diagonal elements $R_{ji}$ are the transition rates from
state $i$ to state $j$. As a result of the evolution equation (\ref{eq:evoeqn}),
the probability distribution of the combined system at the end of
the current interval reads $\boldsymbol{p}_{\tau}(t_{N})=e^{\boldsymbol{R}\tau}\boldsymbol{p}(t_{N})=\boldsymbol{e^{\boldsymbol{R}\tau}}\mathcal{M}\bm{p}^{D}(t_{N}).$
Then the corresponding statistical state of the demon can be written
as 
\[
\bm{p}_{\text{in}}^{D}(t_{N+1})=\mathcal{T}\bm{p}_{\text{in}}^{D}(t_{N}),\ \ \mathcal{T}\equiv\mathcal{P}^{D}e^{\boldsymbol{R}\tau}\mathcal{M}
\]
. According to the Perron-Frobenius theorem \cite{Matrix2000}, any
initial distribution $\bm{p}_{\text{in},0}^{D}$ at the start of the
first interval will evolve asymptotically to a unique periodic steady
state
\begin{equation}
\bm{p}_{\text{in}}^{D,ps}=\lim_{n\rightarrow\infty}\mathcal{T}^{n}\bm{p}_{\text{\text{in},0}}^{D},
\end{equation}
which can be obtained by solving the eigenvalue problem 
\begin{equation}
\mathcal{T}\bm{p}_{\text{in}}^{D,ps}=\bm{p}_{\text{in}}^{D,ps}.
\end{equation}
This unique periodic steady state is just the functional state of
the autonomous Maxwell's demon which can produce anomalous work stably.
To calculate the work produced by the information engine through exploiting
the memory tape, Mandal and Jarzynski have defined a quantity named
as average production:
\begin{equation}
\Phi(\tau)\equiv p_{1}^{f}-p_{1}=p_{0}-p_{0}^{f},
\end{equation}
where $p_{1}^{f}$ and $p_{0}^{f}$ are probabilities of the outgoing
bit to be in states $1$ and $0$ (in the periodic steady state).
The values of $p_{1}^{f}$ and $p_{0}^{f}$ are determined by 
\begin{equation}
\bm{p}_{\tau}^{B,ps}\equiv\left(\begin{array}{c}
p_{0}^{f}\\
p_{1}^{f}
\end{array}\right)=\mathcal{P}^{B}e^{\boldsymbol{R}\tau}\mathcal{M}\bm{p}_{\text{in}}^{D,ps}.
\end{equation}
then the average output work per interaction interval can be computed
as (in the unit of $k_{B}T$)
\[
\langle W\rangle=\Phi(\tau)\cdot w,
\]
with $w$ being the work done by the combined system when a single
jump of the bit from $0$ to $1$ happens. A positive value of $\langle W\rangle$
implies that the autonomous demon is converting information resources
into work. 

\begin{figure}
\begin{centering}
\includegraphics[width=0.8\columnwidth]{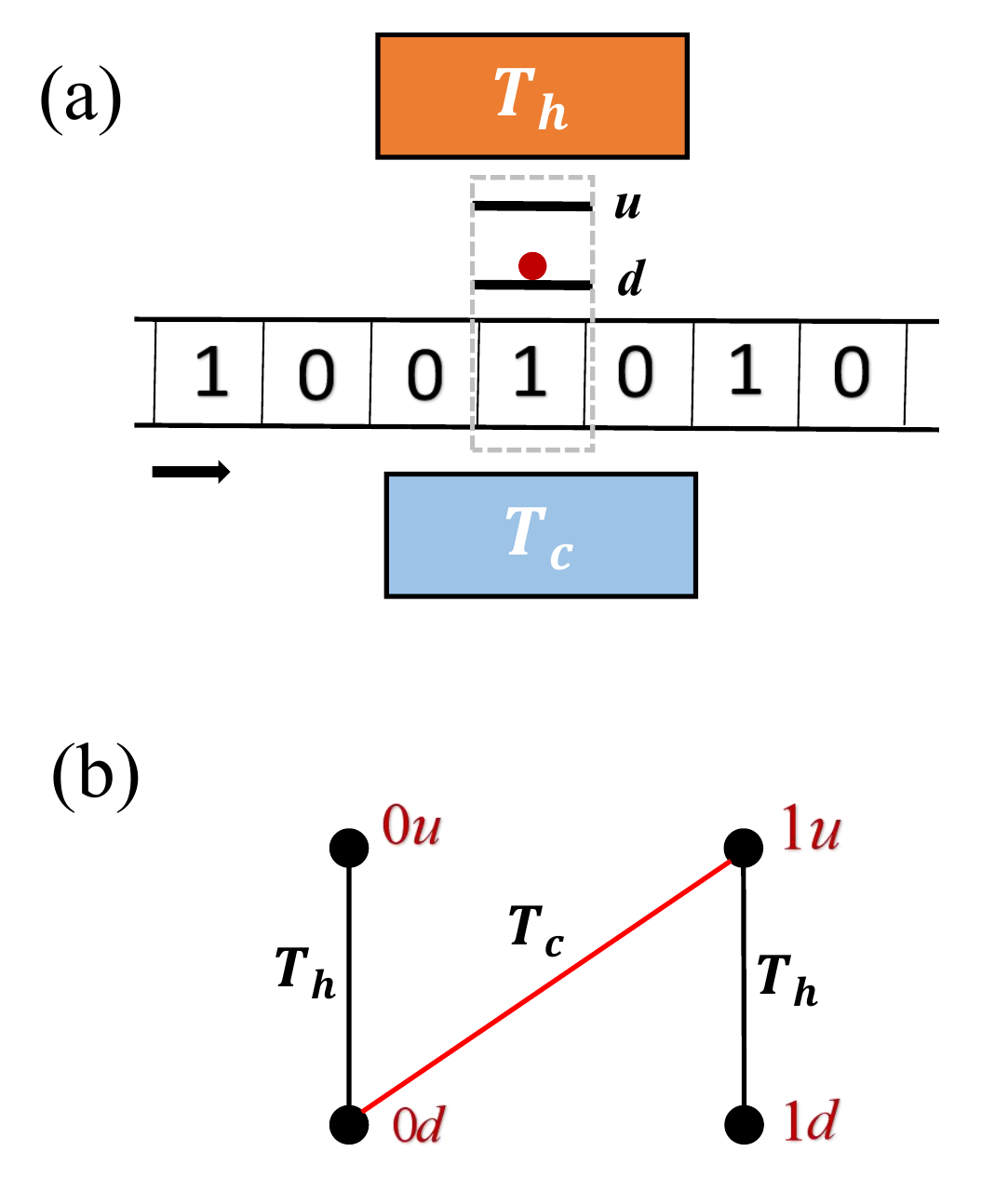}
\par\end{centering}
\caption{The information refrigerator model. (a) The two-state demon interacts
with a sequence of bits and two reservoir at different temperature.
(b) \textcolor{black}{Graph depiction of the composite $4$-state
system. Nodes denote the combined states and edges denote the allowed
transitions, and each pair of transitions satisfies detailed balance
conditions. The red edge represents the transition under the low temperature
$T_{c}$, the other edges correspond to the transition at temperature
$T_{h}$.}}

\label{fig:2}
\end{figure}

We take the simplest two-state demon model (named as information refrigerator)
as an illustrative example, which is our main focus in this work.
A demon with two energy states (up state $u$ with energy $E_{u}$
and down state $d$ with energy $E_{d}$) is coupled with a memory
tape to comprise a four-state combined system (Fig. \ref{fig:2}),
with two heat baths at different temperatures being the environment.
During each interaction interval, the combined system interacts with
the heat bath at high temperature $T_{h}$ when the demon is jumping
randomly between up state and down state by itself, with the current
bit being unchanged. These intrinsic transitions of the demon is irrelevant
to bits. Moreover, another kind of cooperative transitions $0d\leftrightarrow1u$
are allowed , i.e., demon's transitions from down state to up state
shall only occur if the current bit makes a transition from $0$ to
$1$ meanwhile, and from up state to down state only when the bit
transit from $1$ to $0.$ These cooperative transitions happen in
contact with the heat bath at low temperature $T_{c}$, being accompanied
by exchanging energy with the cold heat bath. All transition rates
of intrinsic transitions ($R_{d\rightarrow u},\ R_{u\rightarrow d}$)
and cooperative transitions ($R_{0d\rightarrow1u},\ R_{1u\rightarrow0d}$)
satisfy the detailed balance conditions as follow:
\[
\text{\ensuremath{\frac{R_{d\to u}}{R_{u\to d}}=}}e^{-\beta_{h}\Delta E},\ \text{\ensuremath{\frac{R_{0d\to1u}}{R_{1u\to0d}}=}}e^{-\beta_{c}\Delta E}
\]
where $\beta_{h,c}=1/(k_{B}T_{h,c})$. For later convenience, we parameterize
them as : $R_{d\to u}=\Gamma(1-\sigma),R_{u\to d}=\Gamma(1+\sigma),\ R_{0d\to1u}=1-\omega,R_{u\to d}=1+\omega$.
The characteristic transition rate $\Gamma$ is set to be $1$ in
the rest of the text. Here, $0<\sigma=\tanh\left(\beta_{h}\Delta E/2\right)<1$
and $0<\omega=\tanh\left(\beta_{c}\Delta E/2\right)<1$. We also define
\[
\epsilon=\frac{\omega-\sigma}{1-\omega\sigma}=\tanh\frac{\left(\beta_{c}-\beta_{h}\right)\Delta E}{2},
\]
with $0<\epsilon<1$ quantifying the temperature difference between
the two reservoirs.

For each interaction interval, if a single bit turns from $0$ to
$1$ due to the cooperative transition, then a fixed amount of energy
$\Delta E=E_{u}-E_{d}$ is extracted from the cold reservoir, which
can be identified as the anomalous work done by the demon. Consequently,
the average output work due to the two-state demon in this model reads
$\langle W\rangle\equiv Q_{c\rightarrow h}=\Phi(\tau)\cdot w=\Phi(\tau)(E_{u}-E_{d})$,
where the average production $\Phi(\tau)$ can be computed using equations
(3) - (5). More details about this model are described in the Appendix
B. 

\paragraph*{Formalism of discrete-time stochastic resetting }

Here we would like to introduce a discrete-time resetting mechanism,
which is randomly imposed on the $k$-state demon at the end of each
interaction interval with a probability $\gamma=1-e^{-r\tau}$. Under
this setting, larger the resetting rate $r$ and the time interval
$\tau,$ more possible a resetting event will happen at the end of
an interval. When a resetting event takes place, the demon would be
taken to a given state $\vec{\Delta}$ instantaneously. This discrete-time
resetting process may be realized as follow: resetting events happen
according to a continuous-time Poisson process with rate $r_{0}$,
and all events would only occur very near the end point of each interval,
bringing the demon to the given resetting state $\vec{\Delta}$. Denoting
$\delta t$ as the time window that resetting events can happen near
the end of each interval, the probability that there is at least one
such event is given by $\gamma=1-e^{-r_{0}\delta t}$. We assume that
$\delta t\ll\tau$ and $\delta t\propto\tau$, and define a (modified)
resetting rate $r:=r_{0}\delta t/\tau$. Then the effect of resetting
events near the end of an interval is approximately equal to the effect
of a single resetting event exactly at the end of the interval with
probability $\gamma=1-e^{-r\tau}.$ $\bm{p}_{\text{in,0}}^{D}$ denotes
the initial distribution of demon, which is prepared as an equilibrium
state by letting the demon be in contact with a thermal reservoir
whose temperature is $T_{\text{in}}$ . As mentioned above, in the
absence of resetting, the evolution of the state of demon at the beginning
of each interval can be described as 
\begin{equation}
\bm{p}_{\text{in}}^{D}(t_{N})=\mathcal{T}^{N}\bm{p}_{\text{in,0}}^{D}.\label{eq:original}
\end{equation}
 With fixed-rate resetting events all happening at the end of interaction
intervals, the demon's evolution reads

\begin{equation}
\bm{p}_{\text{in}}^{D}(r,t_{N})=e^{-r\tau}\mathcal{T}\bm{p}_{\text{in}}^{D}(r,t_{N-1})+(1-e^{-r\tau})\vec{\Delta}.\label{eq:r_dynamics}
\end{equation}
Thus the demon's initial distribution at time $t=N\tau$ is formally
determined by 
\begin{equation}
\bm{p}_{\text{in}}^{D}(r,t_{N})=e^{-rN\tau}\bm{p}_{\text{in}}^{D}(t_{N})+\left(1-e^{-r\tau}\right)\sum_{n=0}^{N-1}\left[e^{-rn\tau}\vec{\Delta}_{n\tau}\right],\label{eq:renewal}
\end{equation}
where $\vec{\Delta}_{n\tau}\equiv\mathcal{T}^{n}\vec{\Delta}$ refers
to the state $\vec{\Delta}$ evolves to after $n$ intervals. This
is a renewal equation connecting the distribution at the beginning
of each interval under discrete-time resetting with the distribution
under the reset-free dynamics. On the right hand side of Eq. \ref{eq:renewal},
the first term $e^{-rN\tau}\bm{p}_{\text{in}}^{D}(t_{N})$ accounts
for the situation when there is no resetting events until time $t=N\tau$,
the corresponding probability of which is $e^{-rN\tau}$. The $n^{th}$
term in the summation denotes the case in which the last restart happened
at time $t=(N-n)\tau$, whose probability is $\left(1-e^{-r\tau}\right)e^{-rn\tau}$.
The summation of the probabilities of all possible events mentioned
above is 
\begin{align*}
p_{\text{tot}} & =e^{-rN\tau}+\left(1-e^{-r\tau}\right)\sum_{n=0}^{N-1}e^{-rn\tau}\\
 & =e^{-rN\tau}+1-e^{-rN\tau}=1,
\end{align*}
satisfying the normalization condition. Note that when $\tau\rightarrow0$,
our renewal equation for discrete-time-resetting distribution reduces
to the continuous time counterpart as in \cite{Busiello_2021}.:
\begin{equation}
\bm{p}^{D}(r,t)=e^{-rt}\bm{p}^{D}(t)+r\int_{0}^{t}dt'e^{-rt'}\vec{\Delta}_{t'},
\end{equation}
because when $\tau\rightarrow0$, one has $\left(1-e^{-r\tau}\right)\rightarrow r\tau\equiv rdt'$. 

To obtain the whole dynamics with resetting, it would be helpful
to utilize the spectral analysis method, solving the eigenvalues problem
of the evolution matrix $\mathcal{T}$. The matrix $\mathcal{T}$
has right eigenvectors $\left\{ \bm{R}_{i}\right\} $ and left eigenvectors
$\left\{ \bm{L}_{i}\right\} $ satisfy $\mathcal{T}\bm{R}_{i}=\lambda_{i}\bm{R}_{i}$
and $\bm{L}_{i}^{\text{T}}\mathcal{\mathcal{T}}=\lambda_{i}\bm{L}_{i}^{\text{T}}$,
with $\lambda_{i}$ the eigenvalues, which are sorted as $1=\lambda_{1}>\left|\lambda_{2}\right|\ge\left|\lambda_{3}\right|\ge...$
(we assume that $\text{\ensuremath{\lambda}}_{2}$ is non-degenerate).
Then according to completeness relation the initial state $\bm{p}_{\text{in},0}^{D}$
and resetting state $\vec{\Delta}$ can be expanded separately as
\begin{equation}
\begin{cases}
\bm{p}_{\text{in,0}}^{D} & =\bm{p}_{\text{in}}^{D,ps}+\sum_{i\geq2}a_{i}\bm{R}_{i},\\
\vec{\Delta} & =\bm{p}_{\text{in}}^{D,ps}+\sum_{i\geq2}d_{i}\bm{R}_{i}\text{.}
\end{cases}
\end{equation}
where $a_{i}=\frac{\bm{L}_{i}^{\text{T}}\cdot\bm{p}_{\text{in,0}}^{D}}{\bm{L}_{i}^{\text{T}}\cdot\bm{R}_{i}}$
 and $d_{i}=\frac{\bm{L}_{i}^{\text{T}}\cdot\vec{\Delta}}{\bm{L}_{i}^{\text{T}}\cdot\bm{R}_{i}}$
are coefficients. Thus the state of the demon at the beginning of
the $N^{th}$ time interval can be written as
\begin{equation}
\bm{p}_{\text{in}}^{D}(t_{N})=\mathcal{T}^{N}\bm{p}_{\text{in,0}}^{D}\bm{=p}_{\text{in}}^{D,ps}+\sum_{i\geq2}a_{i}\lambda_{i}^{n}\bm{R}_{i},\label{eq:N_init}
\end{equation}
and 
\begin{equation}
\vec{\Delta}_{n\tau}\equiv\mathcal{T}^{n}\vec{\Delta}=\bm{p}_{\text{in}}^{D,ps}+\sum_{i\geq2}d_{i}\lambda_{i}^{n}\bm{R}_{i}\text{.}\label{eq:n_reset}
\end{equation}
One can identify the second term on the right hand side of Eq. (\ref{eq:N_init})
as a slowest decaying mode dominating the relaxation time scale, once
the second coefficient $a_{2}$ is not equal to zero. In this case,
the relaxation timescale is typically characterized by $\tau_{\text{rel}}=-1/\ln\left|\lambda_{2}\right|$. 

\begin{figure}
\begin{centering}
\includegraphics[width=1.0\columnwidth]{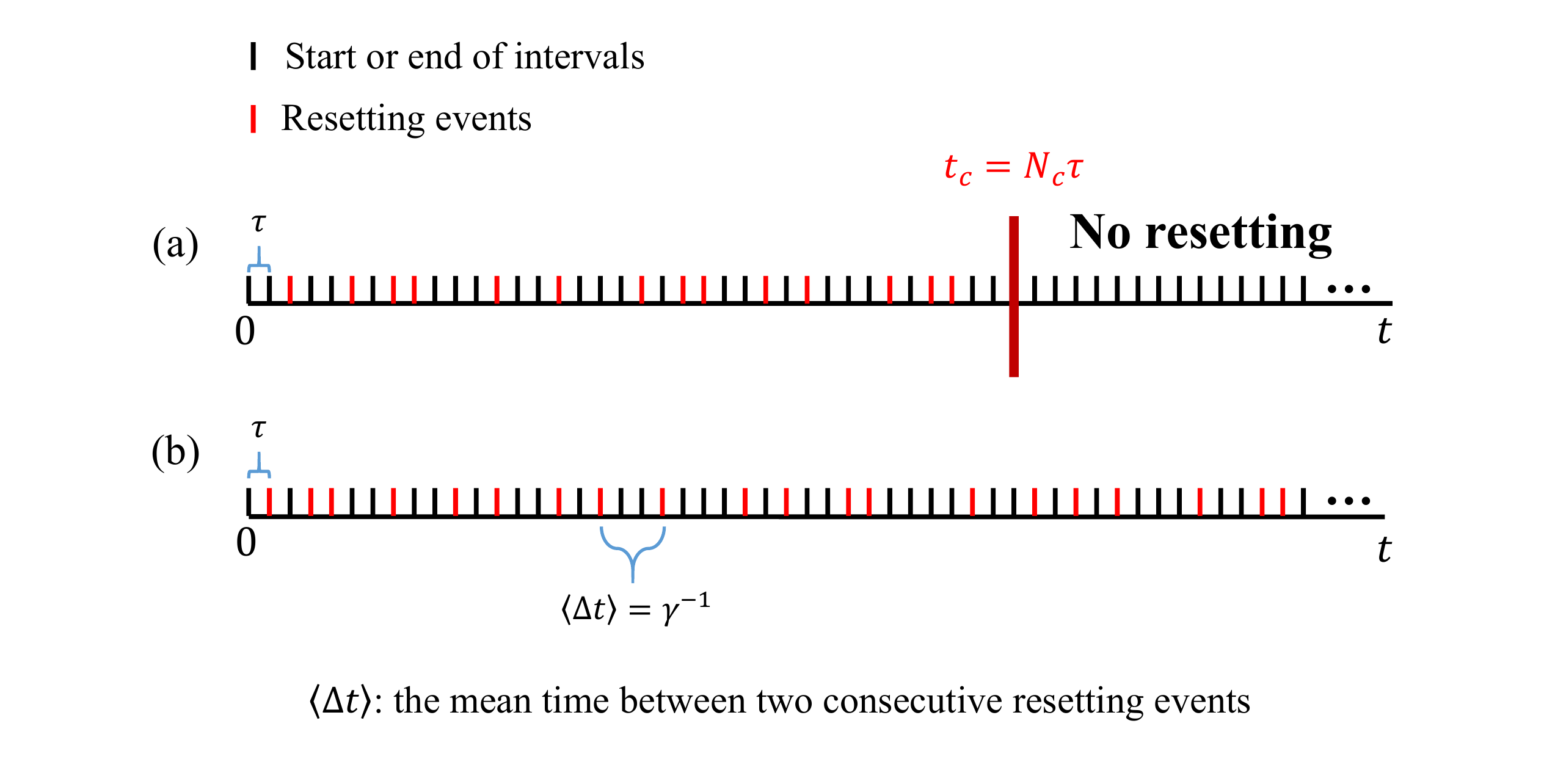}
\par\end{centering}
\caption{An illustration of two resetting strategies. (a) The first resetting
strategy: closing the reset after a critical time $t_{c}$ (b) The
second resetting strategy: always keeping the reset on.}

\label{fig:2-1}
\end{figure}

Next, we gives some detailed descriptions of the two resetting strategies
we would use to devise the autonomous demon. The first resetting strategy
is randomly resetting the demon to the reset state and then switching
off the resetting after a given time $t_{c}=N_{c}\tau$, causing the
system to evolve according to the original dynamics without resetting.
The whole dynamics of the first strategy can be formulated as 
\begin{align}
\bm{p}_{\text{in}}^{D}(r,t_{N})= & \left[\gamma\mathcal{T}\bm{p}_{\text{in}}^{D}(r,t_{N-1})+(1-\gamma)\vec{\Delta}\right]\Theta(N_{c}-N)\nonumber \\
+ & \mathcal{T}\bm{p}_{\text{in}}^{D}(r,t_{N-1})\Theta(N-N_{c})\text{,}
\end{align}
where $\Theta$ is the Heaviside step function. The first strategy
may used to eliminate the slowest decaying mode (making $a_{2}=0$)
so that the demon would reach its functional state at a greatly faster
pace. The second strategy is simply to keep the stochastic resetting
mechanism always on so that the combined system will eventually reach
a new periodic steady state, whose properties depend on the resetting
rate $r$. For an illustration of the two resetting strategies, see
Fig. \ref{fig:2-1}.

Note that the resetting state is chosen to be a single state {[}e.g.,
$(0,1)^{\text{T}}$ for the two-state demon{]} instead of a mixed
state in this work. Physically, resetting the demon to a mixture of
distinct states effectively equals to resetting it to different single
states randomly with different probabilities, which may be more technical
for experimental realizations. Equipped with these discrete-time resetting
formalism and the eigenvector-expansion formula, the designing principles
for autonomous Maxwell's demons are provided in the next two sections.

\section{Inducing Faster Relaxation}

In this section, we show that how the first resetting strategy can
accelerate the relaxation from an arbitrary initial distribution to
demon's functional steady state significantly. This significantly
fast relaxation phenomenon induced by the stochastic resetting is
similar to the Markovian Mpemba effect \cite{lu2017nonequilibrium,klich2019mpemba,santos2020mpemba,baity2019mpemba}.
We just turn off the resetting mechanism at a critical time $t_{c}=N_{c}\tau$,
after which the demon system obeys the original evolutionary dynamics
(\ref{eq:original}) with the coefficient of the relaxation mode ($i^{th}$
coefficient of the eigenvector expansion) $\bm{R}_{i}$ being $a_{i}(r,N_{c})$.
Plugging (\ref{eq:N_init}) and (\ref{eq:n_reset}) into equation
(\ref{eq:renewal}) one can obtain (see Appendix A for details)
\begin{equation}
\bm{p}_{\text{\text{in}}}^{D}(r,t_{N})=\bm{p}_{\text{in}}^{D,ps}+\sum_{i\geq2}a_{i}(r,N)\lambda_{i}^{N}\bm{R}_{i},
\end{equation}
where the modified $i^{th}$ coefficient at $N^{th}$ interaction
interval $a_{i}(r,N)$ reads
\begin{equation}
a_{i}(r,N)=\left[a_{i}-\frac{d_{i}\left(1-e^{-r\tau}\right)}{1-\lambda_{i}e^{-r\tau}}\right]e^{-rN\tau}+\frac{d_{i}\left(1-e^{-r\tau}\right)}{1-\lambda_{i}e^{-r\tau}}\cdot\lambda_{i}^{-N}.\label{eq:a2}
\end{equation}
Here $d_{i}$ is the $i^{th}$ coefficient of the expanded form of
the $\vec{\Delta}$, depending on the choice of the reset state. Therefore,
the whole dynamics under the first resetting strategy obeys

\begin{equation}
\begin{cases}
\bm{p}_{\text{\text{in}}}^{D}(r,t_{N})=\bm{p}_{\text{in}}^{D,ps}+\sum_{i\geq2}a_{i}(r,N)\lambda_{i}^{N}\bm{R}_{i} & N\leq N_{c}\\
\bm{p}_{\text{\text{in}}}^{D}(r,t_{N})=\bm{p}_{\text{in}}^{D,ps}+\sum_{i\geq2}a_{i}(r,N_{c})\lambda_{i}^{N}\bm{R}_{i} & N>N_{c}
\end{cases}.
\end{equation}
By closing the reset at an appropriate critical time $t_{c}$, one
can make the second coefficient being zero so that the relaxation
gets significantly faster, corresponding to the so-called ``strong''
Mpemba effect. Combining $a_{2}(r,N)=0$ with Eq. (\ref{eq:a2}) one
gets the appropriate critical number of interaction intervals if the
second eigenvalue $\lambda_{2}>0$ ($\lambda_{2}$ is demonstrated
to be always positive in the models we study):

\begin{equation}
N_{c}=\frac{1}{r\tau-\ln\lambda_{2}}\ln\left[1-\frac{a_{2}}{d_{2}}\frac{1-\lambda_{2}e^{-r\tau}}{1-e^{-r\tau}}\right],\label{eq:result1}
\end{equation}
which is our first main result. From the expression of $N_{c}$ we
clearly see that the condition for the existence of a physical critical
number $N_{c}\geq1$ is just $a_{2}/d_{2}\leq0$. Furthermore, we
are able to let $N_{c}$ be a small number like one through modifying
the resetting rate $r$ whenever the condition $a_{2}/d_{2}\leq0$
is fulfilled, thus the resetting strategy could always significantly
reduce the total time cost to demon's functional state compared to
the reset-free dynamics. To show how this can be utilized to improve
the performance of autonomous Maxwell's demon, we focus on a specific
model, the two-state information refrigerator mentioned above. 

\begin{figure}
\begin{centering}
\includegraphics[width=0.8\columnwidth]{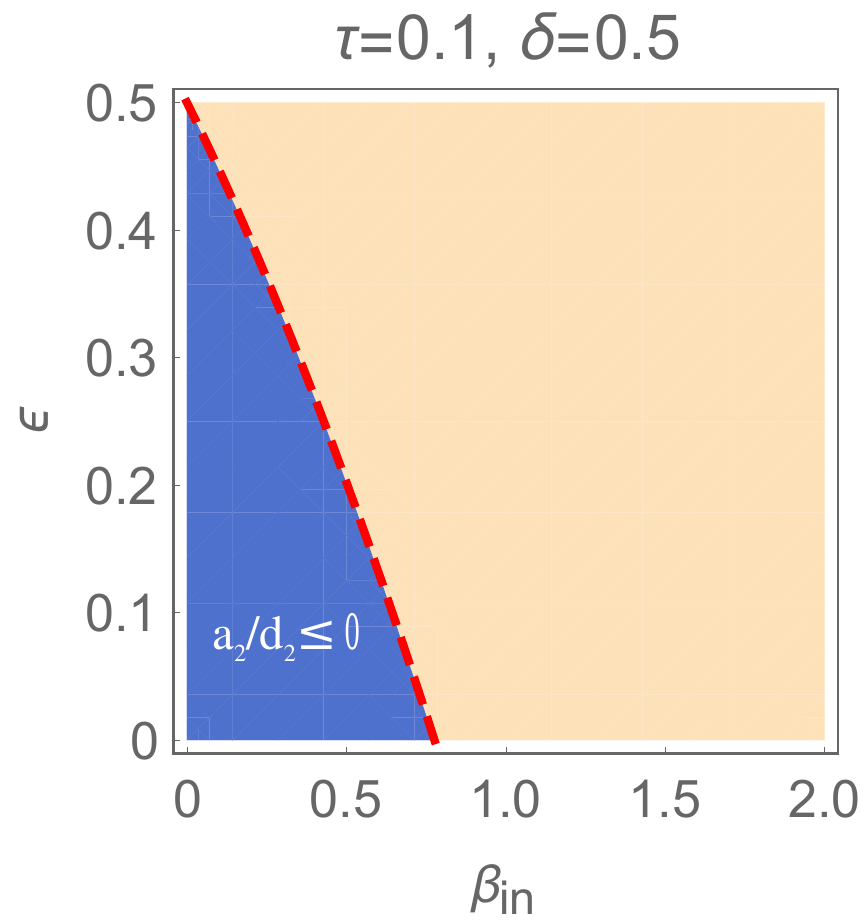}
\par\end{centering}
\caption{The phase diagram for the relaxation behavior of the information refrigerator.
The parameter is set as $\tau=0.1,\ $$\delta=0.5,$ $\Delta E=1$
and the reset state is the down state. The blue are is the fast relaxation
region ($a_{2}/d_{2}\protect\leq0$), where the demon can reach the
functional state significantly faster by stochastic resetting with
appropriate rate. The yellow part is the dub region where resetting
cannot take effect. However, if the reset state is set to be the up
state, the yellow region would turn to be the useful region instead.}

\label{fig:1}
\end{figure}

For the two-state information refrigerator, the resetting state could
be $\vec{\Delta}_{d}=(0,1)^{\text{T}}$ or $\vec{\Delta}_{u}=(1,0)^{\text{T}}$,
we choose the former one. Initially, we let the two-state demon be
in contact with a heat bath whose temperature is $T_{\text{in}}$
for long enough time so that the demon reaches the thermal equilibrium.
Then the initial distribution of the two-state demon is just 
\begin{equation}
\bm{p}_{\text{in,0}}^{D}=\left(\frac{e^{-\Delta E/T_{\text{in}}}}{1+e^{-\Delta E/T_{\text{in}}}},\frac{1}{1+e^{-\Delta E/T_{\text{in}}}}\right)^{\text{T}},
\end{equation}
which makes the second coefficient $a_{2}(T_{\text{in}})$ be a function
of the initial temperature, so is the modified second coefficient
$a_{2}(r,N,T_{\text{in}})$. It is clear from Eq. (\ref{eq:result1})
that once the initial temperature allows the critical interaction
number to be positive, then one can always make $N_{c}$ the smallest
positive integer $1$ by virtue of adjusting the resetting constant
$r$, whatever the dynamical details (i.e. values of $\delta=p_{0}-p_{1}$
and the temperature difference quantifier $\epsilon$ of the two heat
baths) of the system is. Note that for the information refrigerator
system, there is only one right eigenvector $\bm{R}_{2}$ as the relaxation
mode, i.e.
\[
\bm{p}^{D}(r,t_{N})=\bm{p}_{\text{in}}^{D,ps}+a_{2}(r,N,T_{\text{in}})\lambda_{2}^{N}\bm{R}_{2}.
\]
In result, $a_{2}(r,N_{c},T_{\text{in}})=0$ signifies the arrival
of the functional periodic steady state $\bm{p}_{\text{in}}^{D,ps}$.
Thus for the two-state demon one can invariably expect the shortest
time $t_{c}=\tau$ to enter the functional state through controlling
the resetting parameter $r.$ In Fig. \ref{fig:1} we build a phase
diagram for the dynamical behavior of the information engine before
reaching its functional state. The blue area of the phase diagram
corresponds to the fast relaxation ($a_{2}/d_{2}\leq0$) region, where
an optimal resetting rate that can lead to the smallest time cost
$t_{c}=\tau$ always exists. This diagram provides guidance to prepare
the initial state of the demon to be in an appropriate temperature.
What's more, if we set the up state as the reset state, the yellow
area in the current diagram would turn to be the efficacious region
of resetting. The asymmetry of the efficacious region corresponding
to up state and down state arises from the energy difference between
two states. 

\begin{figure}
\begin{centering}
\includegraphics[width=0.8\columnwidth]{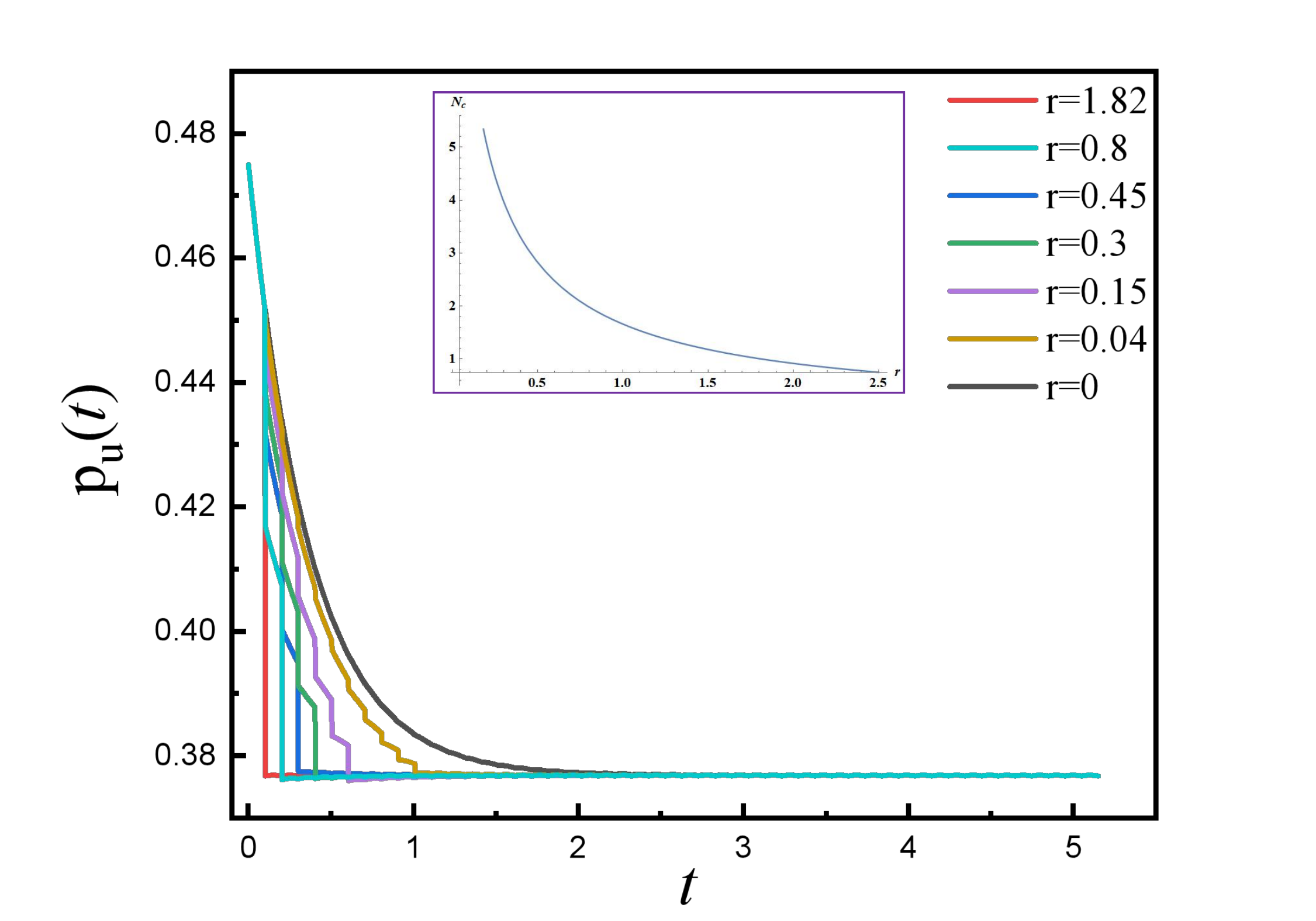}
\par\end{centering}
\caption{The evolution of the demon's state under different resetting dynamics.
For each dynamics with different resetting rate $r$, the resetting
is closed right after the critical time $t_{c}(r)=N_{c}(r)\tau.$
We take the dynamics of probability of the up state as an illustration.
The dynamical parameters are given by $\tau=0.1,\ \delta=0.5,\ \epsilon=0.2$
and $\beta_{\text{in}}=0.1$. The inset shows the relation between
the critical number $N_{c}$ and the resetting rate $r$.}

\label{fig:1-1}
\end{figure}

Then, to illustrate the accelerating effect of stochastic resetting,
we further fix the dynamical parameters as $\delta=0.5,\ \epsilon=0.2$
and prepare the demon initially at $\beta_{\text{in}}=0.1$, then
numerically gives several dynamical trajectories of the probability
of the up state $p_{u}(t)$ under different resetting rates, as plotted
in Fig. \ref{fig:1-1}. Note that the resetting dynamics is switched
off after a given critical time $t_{c}(r)=N_{c}(r)\tau$, depending
on the resetting rate $r$. The optimal value of $r$ which can make
$t_{c}=\tau$ is also a threshold value. It is shown in Fig. \ref{fig:1-1}
that when the resetting rate $r$ is below this threshold value, the
larger the $r$ is, the sooner the refrigerator could get to its functional
periodic steady state. However, when the resetting rate is made to
be larger than the optimal value, the time cost would become greater
than $\tau.$ In the limiting case of $r\rightarrow\infty$, the demon
would be reset to the down state (which can never be the functional
state) with probability $1$ at the end of each interval, making the
critical time $t_{c}$ go to zero, which is the same as the reset-free
dynamics. In this given specific case, the threshold value of $r$
(or the optimal $r$) turns to be $1.82.$ To make it clearer, we
plot the relation between the critical number $N_{c}$ and resetting
rate $r$ in the inset of Fig.\ref{fig:1-1}, with the parameters
being fixed at the same values as above.

\section{Autonomous Demon with Resetting }

In this section, the second resetting strategy is taken into consideration,
and the second features of the demon's performance, i.e., the efficacious
working region, is studied. To study the efficacious working region
of demon, we first need to analyze the functional state performance
of the autonomous demon with the second strategy being imposed on
it. There are two crucial aspects of the functional state performance
of the autonomous demon, including the average work production $\Phi$
per cycle and the information erasing capability quantified by the
decrease of Shannon entropy of bit per cycle. If the autonomous demon
is always under resetting, it will finally reach a new periodic steady
state in the large time limit (the number of interactions $N\rightarrow\infty$),
which is given by (see Appendix A for derivations)
\begin{equation}
\bm{p}_{\text{in}}^{D,ps}(r)=\bm{p}_{\text{in}}^{D,ps}+\sum_{i\geq2}\frac{1-e^{-r\tau}}{1-\lambda_{i}e^{-r\tau}}d_{i}\bm{R}_{i}.\label{eq:new_pss}
\end{equation}
As a consequence, the marginal distribution of the outgoing bit in
the new periodic steady state can be written as 
\begin{equation}
\mathcal{P}_{\tau}^{B,ps}(r)=\mathcal{P}^{B}e^{\mathcal{R}\tau}\mathcal{M}\bm{p}_{\text{in}}^{D,ps}(r)=(p_{0}^{r},p_{1}^{r})^{\text{T}}.
\end{equation}
Then, we would like to quantify the two important properties mentioned
above in this new functional state, i.e., the average work production
and the information erasing capability of the demon, so that the efficacious
region could be analyzed in detail. Therefore, in the remaining part
of this section, we assume the demon has reached its new functional
periodic steady state. For simplicity, we take the two-state Maxwell's
refrigerator model as an illustrative example. The distribution of
demon at the start of each interval is 
\[
\bm{p}_{\text{in}}^{D,ps}(r)=\bm{p}_{\text{in}}^{D,ps}+\frac{1-e^{-r\tau}}{1-\lambda_{2}e^{-r\tau}}d_{2}\bm{R}_{2}.
\]
The performance of the information refrigerator with resetting can
be evaluated by the average production of 1's per interaction interval,
recalling that the initial distribution of bit is given by $\bm{p}_{\text{\text{in}}}^{B}=(p_{0},p_{1})^{T}$:
\begin{equation}
\Phi_{\text{tot}}(r,\tau)\equiv p_{1}^{r}-p_{1}=p_{0}-p_{0}^{r}.
\end{equation}
The average transfer of energy from the cold to the hot reservoir
is $Q_{c\rightarrow h}=\Phi_{r}\Delta E=\Phi_{r}(E_{u}-E_{d}).$ The
total average production per interaction interval is given by
\begin{align}
\Phi_{\text{tot}} & =\left(\left[\mathcal{T}^{'}\bm{p}_{\text{in}}^{D,ps}\right]_{2}-p_{1}\right)+\frac{1-e^{-r\tau}}{1-\lambda_{2}e^{-r\tau}}d_{2}\left[\mathcal{T}^{'}\bm{R}_{2}\right]_{2},\nonumber \\
 & \equiv\Phi_{0}+\Phi_{r}
\end{align}
where $\mathcal{T}^{'}=\mathcal{P}^{B}e^{\mathcal{R}\tau}\mathcal{M}$.
It has been shown that (see Appendix B for details)
\begin{equation}
\Phi\equiv\left[\mathcal{T}^{'}\bm{p}_{\text{in}}^{D,ps}\right]_{2}-p_{1}=\frac{\delta-\epsilon}{2}\eta(\Lambda),
\end{equation}
thus we just need to compute the contribution $\Phi_{r}$ arising
from stochastic reset. The second right eigenvector is obtained as
$\boldsymbol{R}_{2}=(1,-1)^{T}$. We set the resetting state as $\vec{\Delta}=(0,1)^{T}$(down
state) or $\vec{\Delta}=(1,0)^{T}$ (up state), and study their contributions
respectively. When the demon is reset to the up state, we find that
the contribution $\Phi_{r}$ always takes negative value whatever
the value of $r$ is. This implies that the up state is not a good
choice for the reset state because resetting the demon to up state
only has negative impact on the engine's performance, i.e., shrinking
the refrigerator region. We will see that the down state $\vec{\Delta}=(0,1)^{T}$
would be an appropriate option for the demon to be reset to, even
might be optimal. 

Now consider another important feature of performance of the refrigerator
in the new functional state, the information-processing capability
of the demon. To quantify the capability, the Shannon entropy difference
between the outgoing be and the incoming bit is introduced as
\begin{align}
\Delta S_{B}= & S(\boldsymbol{p}_{\text{\ensuremath{\tau}}}^{B,ps})-S(\boldsymbol{p}^{B,ps})=S(\delta-2\Phi_{\text{tot}})-S(\delta),\nonumber \\
S(\delta)\equiv & -\sum_{i=1}^{1}p_{i}\ln p_{i}\nonumber \\
= & -\frac{1-\delta}{2}\ln\frac{1-\delta}{2}-\frac{1+\delta}{2}\ln\frac{1+\delta}{2},
\end{align}
which is a measure of how much information content contained in the
memory tape is changed due to the demon during each interaction interval.

\begin{figure}
\begin{centering}
\includegraphics[width=0.8\columnwidth]{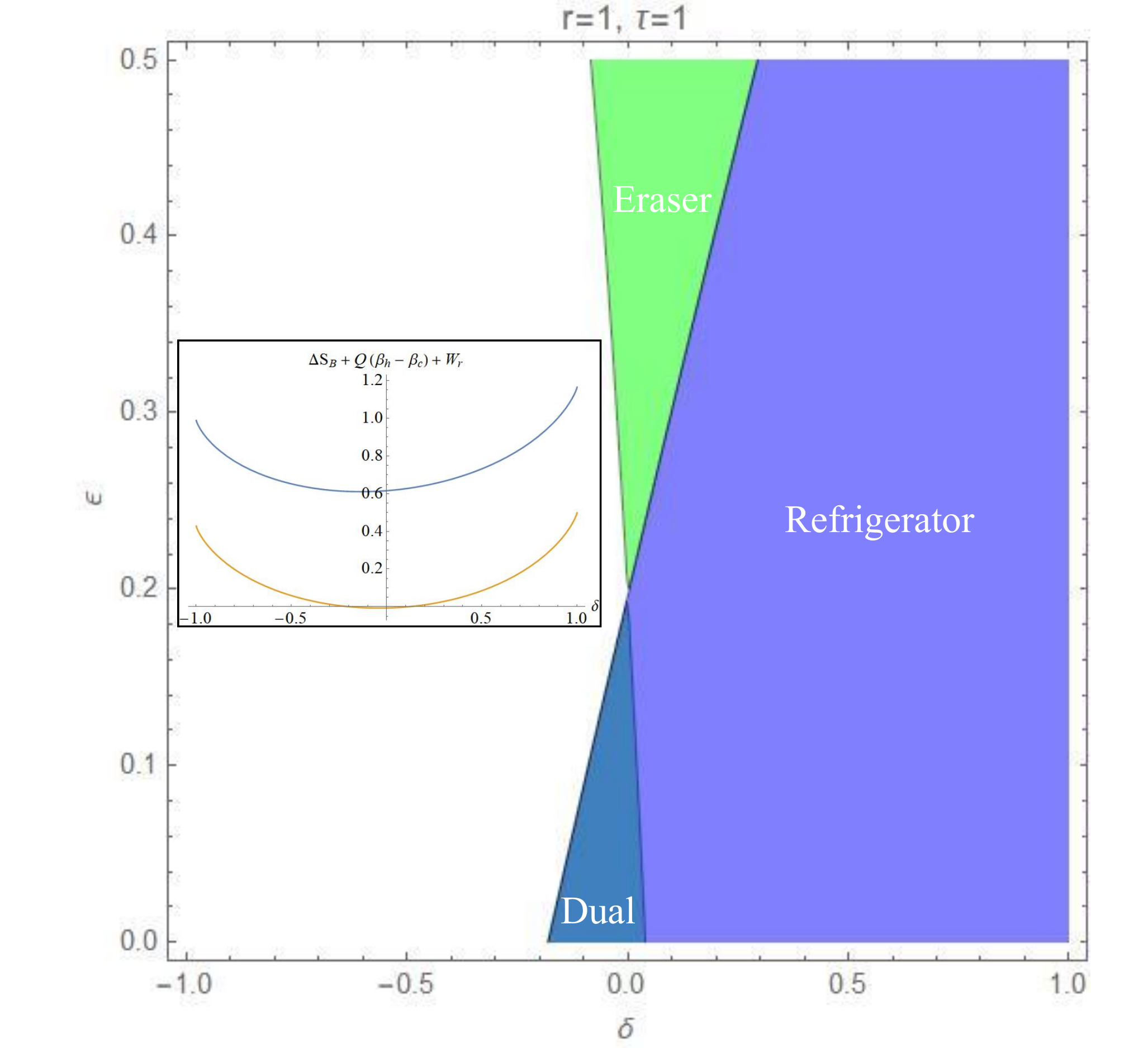}
\par\end{centering}
\caption{The new phase diagram. resetting rate $r=1$, interaction interval
$\tau=1$. The green region is the information-eraser region, in which
the demon is able to reduce the information entropy of the memory
tape. The purple region is the refrigerator region, where the demon
can help to produce anomalous work by exploiting information. The
blue region is a dual-function region, both of the information erasing
and anomalous work production functions could be realized in this
region. The white part is the dub region. The inset shows a demonstration of the generalized second law, where the dynamical parameter $\epsilon$ is fixed to be $0.06$. The yellow line corresponds to the line of original second law without considering the contribution from resetting, the blue line corresponds to the generalized second law including the resetting work term $W_r$.}

\label{fig:4}
\end{figure}

At the end of the day, to illustrate the extended efficacious region
of the refrigerator, we fix $\Gamma=1,\ \omega=1/2$ and construct
a phase diagram for the refrigerator, traversing the parameter space
of $(\delta,\epsilon)$ (Fig. \ref{fig:4}). Note that to assure $\beta_{c}>\beta_{h}>0$,
parameter $\epsilon$ can only range from $0$ to $1/2$, since we
have set $\omega=\tanh(\beta_{c}\Delta E/2)=1/2>\epsilon=\tanh[(\beta_{c}-\beta_{h})\Delta E/2]$.
The new phase diagram is comprised of four different regions, and
it's our second main result. The purple part turns to be the information
refrigerator region, which is obviously extended compared to the original
reset-free demon case because of the positive contribution $\Phi_{r}$
from resetting. The green part is the information eraser region, where
the information encoded in the memory tape could be wiped out, restoring
the low-information-entropy state as a source for anomalous work.
Strikingly, the blue part is a dual-function region in which the information
machine can produce anomalous work and supplement some information
resource meanwhile. 

\paragraph*{Generalized second law in the presence of resetting}

The dual-function region in the new phase diagram shows that the autonomous
demon with resetting seems to be against the second law of thermodynamics.
To address this issue, a generalized second law is derived by constructing
a Lyapunov function, or by using the integral fluctuation theorem
for stochastic entropy production (see Appendix E for details).

As a result, the generalized second law of thermodynamics of the system
with resetting in the periodic steady state is demonstrated to be
\begin{equation}
Q_{c\rightarrow h}(\beta_{h}-\beta_{c})+\Delta S_{B}+\beta_{h}W_{\tau}^{r}\geq0,\label{eq:2ndlaw}
\end{equation}
where the $W_{\tau}^{r}$ is the 'resetting work' during an interval
$[n\tau,(n+1)\tau]$ due to stochastic resetting. This generalized second law is our third main result. The new term $W_{\tau}^{r}$
is the cost for resetting, and for the information refrigerator it
is given by
\begin{equation}
W_{\tau}^{r}=k_{B}T_{h}\Delta S_{D}+\Delta p^{D,ps}\Delta E.\label{eq:reset_cost}
\end{equation}
Here, $\Delta p^{D,ps}\equiv p_{\tau,u}^{D,ps}-p_{\text{in},u}^{D,ps}$
denotes the difference between the steady state probability of the
demon in up state at the final moment and at the initial moment within
an interval. From the expression, we can see that this new term contributed
by resetting consists of two parts. The first part is the Shannon
entropy difference between the initial distribution and the final
distribution in each interaction interval, which is an extra entropic
cost. The second part $\beta_{h}\Delta p^{D,ps}\Delta E$ can be interpreted
as the energy cost needed to maintain this distribution difference
between the demon at the final moment and at the initial moment, or
say, to maintain the new periodic steady state. To demonstrate the generalized second law, we further fix $\epsilon$ to be $0.06$ and plot the $Q_{c\rightarrow h}(\beta_{h}-\beta_{c})+\Delta S_{B}$ and $Q_{c\rightarrow h}(\beta_{h}-\beta_{c})+\Delta S_{B}+\beta_{h}W_{\tau}^{r}$ respectively as functions of $\epsilon$ in the inset of Fig. \ref{fig:4}. The yellow line corresponds to the original second law without considering the effect of resetting, and the blue line corresponds to the generalized second law. It can be seen that part of the yellow line is below zero, which means the original second law is violated. What's more, using
the so-called thermodynamic uncertainty relation (TUR) \cite{TUR2015_Seifert,TUR2016_England,TUR2016_Esposito,TUR2020_arbitrary,TUR2021_Seifert},
a relation stronger than the generalized second law (\ref{eq:2ndlaw})
may be obtained.

From this expression (\ref{eq:reset_cost}) we may explain physically
why the down state $\vec{\Delta}=(0,1)^{\text{T}}$ should be chosen
as the resetting state to improve the engine's performance, instead
of the up state $\vec{\Delta}=(1,0)^{\text{T}}$ or any mixed state.
Actually the down state may be the best selection that can optimize
the performance of the information refrigerator. To better the comprehensive
performance of the information refrigerator, larger cost from resetting
$W_{\tau}^{r}$ is preferred, because the larger the cost is, the
bigger the anomalous energy transfer $Q_{c\rightarrow h}$ and the
smaller the entropy difference $\Delta S_{B}$ can be. Therefore,
the optimal resetting mechanism should maximize the $\Delta p^{D,ps}$
and the $\Delta S_{D}$ simultaneously. If the demon is reset to a
given mixed state like $(1/2,1/2)^{\text{T}}$, the initial Shannon
entropy $S_{D,0}$ would take the maximal value, making $\Delta S_{D}<0$.
Compared to this, reset the demon to a single state like $(0,1)^{\text{T}}$
or $(1,0)^{\text{T}}$ would make the Shannon entropy of the demon
at the initial moment of an interval being zero, always following
with $\Delta S_{D}>0$. Thus resetting the demon to a single state
could be superior to resetting it to a mixed state. On the other hand,
if the demon is reset to the up state at the start of each interval,
the probability of it being in up state initially turns to be $1$,
leading $\Delta p^{D,ps}$ to equals $p_{\tau,u}^{D,ps}-1<0$, which
have negative impact in the refrigerator's performance. Contrarily,
the down-state reset still bring about the positive effect for this
term as $\Delta p^{D,ps}=p_{\tau,u}^{D,ps}>0$. In consequence, picking
the down state $(0,1)^{\text{T}}$ as the reset state rather than
other states seems to be the optimal strategy.

\section{Discussion}

Autonomous Maxwell's demons have been paid much attention to in the
field of small-system thermodynamics since Mandal and Jarzynski declared
their exactly solvable model in 2012. It should be noted that the
evolution of the memory tape is under some kind of resetting mechanism,
i.e. the state of the tape is always being reset to a given initial
statistical state after a fixed time $\tau$. Therefore, it would
be intriguing to let the demon exposed to stochastic resetting as
the bit does, which may serve as a strategy improving the demon's
performances. It has been shown that resetting protocol can be devised
to optimize some dynamical processes \cite{2019Stochastic,2022reset,Busiello_2021},
thus it might be a promising idea to reset the demon randomly. However,
most existing frameworks of stochastic resetting only deal with the
continuous-time Markov process, which aren't applicable to our case.
Because the reset of the demon is only wanted to happen at the end
of some interaction intervals. 

In this article, we generalize the stochastic resetting formalism
to discrete-time Markov process, so that this mechanism can be introduced
to the autonomous Maxwell's demon system, providing some novel designing
principles and strategies. The time cost for the autonomous demon
to relax to its working state (the periodic steady state) hasn't been
taken into account in the previous works. In reality, minimizing this
time cost for the relaxation processes before the working state would
be preferred. Using stochastic resetting mechanism, we provide some
designing principles to optimize this relaxation process, minimizing
the time cost to enter the working state through a protocol inspired
by the strong Mpemba effect. Furthermore, we construct new functional
states with remarkable features by keeping resetting the demon, and
derive a generalized second law of thermodynamics including the contribution
from discrete-time stochastic reset. We have illustrated our designing
strategies in two autonomous demon models, the two-state Maxwell's
refrigerator and the three-state information heat engine. 

An interesting open problem is the generalization of our framework
to the case of time-dependent resetting rate, where the waiting time
distribution of two consecutive resetting events is not an exponential
distribution. To find other waiting time distributions which can improve
the performance of the autonomous demon, it is really worthy of
further study. Moreover, the stochastic resetting in quantum system
is recently of wide interest \cite{22quantumreset,21PRB_reset}, thus
it would be promising to develop a framework as the quantum counterpart
of our discrete-time resetting framework, which may further serve
as a useful tool to analyze resetting mechanism in quantum systems.
\begin{acknowledgments}
This work is supported by MOST (Grant No. 2018YFA0208702), NSFC (Grant Nos. 32090044, 21790350 and 21521001).
\end{acknowledgments}

\appendix
\onecolumngrid

\section*{Appendix}


\section{Modified dynamics under stochastic resetting}

Here we provide some details for the derivation of the equation (\ref{eq:a2})
and (\ref{eq:new_pss}) in the main text.

Plugging the expansion formula 
\[
\bm{p}_{\text{\text{in}}}^{D}(t_{N})=p_{\text{in}}^{D,ps}+\sum_{i\geq2}a_{i}\lambda_{i}^{n}\bm{R}_{i}
\]
 and 
\[
\vec{\Delta}_{n\tau}=\bm{p}_{\text{in}}^{D,ps}+\sum_{i\geq2}d_{i}\lambda_{i}^{n}\bm{R}_{i}
\]
into the renewal equation (\ref{eq:renewal}) for the demon's initial
distribution:

\[
\bm{p}_{\text{in}}^{D}(r,t_{N})=e^{-rN\tau}\bm{p}_{\text{in}}^{D}(t_{N})+\left(1-e^{-r\tau}\right)\sum_{n=0}^{N-1}\left[e^{-rn\tau}\vec{\Delta}_{n\tau}\right],
\]
we reach that
\begin{align}
\bm{p}^{D}(r,t_{N})= & e^{-rN\tau}\left[\bm{p}_{\text{in}}^{D,ps}+\sum_{i\geq2}a_{i}\lambda_{i}^{n}\bm{R}_{i}\right]+\left(1-e^{-r\tau}\right)\sum_{n=0}^{N-1}e^{-rn\tau}\left[\bm{p}_{\text{in}}^{D,ps}+\sum_{i\geq2}d_{i}\lambda_{i}^{n}\bm{R}_{i}\right]\\
= & \left[\bm{p}_{\text{in}}^{D,ps}+\sum_{i\geq2}\frac{1-e^{-r\tau}}{1-\lambda_{i}e^{-r\tau}}d_{i}\bm{R}_{i}\right]+\sum_{i\geq2}\left[a_{i}-\frac{1-e^{-r\tau}}{1-\lambda_{i}e^{-r\tau}}d_{i}\right]e^{-rN\tau}\lambda_{i}^{N}\bm{R}_{i}\label{eq:npss}\\
= & \bm{p}_{\text{in}}^{D,ps}+\sum_{i\geq2}\left\{ \left[a_{i}-\frac{d_{i}\left(1-e^{-r\tau}\right)}{1-\lambda_{i}e^{-r\tau}}\right]e^{-rN\tau}+\frac{d_{i}\left(1-e^{-r\tau}\right)}{1-\lambda_{i}e^{-r\tau}}\cdot\lambda_{i}^{-N}\right\} \lambda_{i}^{N}\bm{R}_{i}\\
\equiv & \bm{p}_{\text{in}}^{D,ps}+\sum_{i\geq2}a_{i}(r,N)\lambda_{i}^{N}\bm{R}_{i}.
\end{align}
Here, the modified coefficients $a_{i}(r,N)$ is obtained as 
\[
a_{i}(r,N)=\left[a_{i}-\frac{d_{i}\left(1-e^{-r\tau}\right)}{1-\lambda_{i}e^{-r\tau}}\right]e^{-rN\tau}+\frac{d_{i}\left(1-e^{-r\tau}\right)}{1-\lambda_{i}e^{-r\tau}}\cdot\lambda_{i}^{-N},
\]
and the Eq. (\ref{eq:npss}) gives rise to the initial distribution
of demon's new periodic steady state as 
\begin{equation}
\bm{p}_{\text{in}}^{D,ps}(r)=\lim_{N\rightarrow\infty}\bm{p}^{D}(r,t_{N})=\bm{p}_{\text{in}}^{D,ps}+\sum_{i\geq2}\frac{1-e^{-r\tau}}{1-\lambda_{i}e^{-r\tau}}d_{i}\bm{R}_{i},
\end{equation}
which is the Eq. (\ref{eq:new_pss}) in the main text.

\section{Details of information refrigerator}

\paragraph*{Extra details of the original information refrigerator}

The transition matrix of the four-state combined system in this model
is given by
\[
\boldsymbol{R}=\begin{pmatrix}-\Gamma(1+\sigma) & \Gamma(1-\sigma) & 0 & 0\\
\Gamma(1+\sigma) & -\left[1-\omega+\Gamma(1-\sigma)\right] & 1+\omega & 0\\
0 & 1-\omega & -\left[1+\omega+\Gamma(1+\sigma)\right] & \Gamma(1-\sigma)\\
0 & 0 & \Gamma(1+\sigma) & -\Gamma(1-\sigma)
\end{pmatrix}
\]

Solution to the periodic steady state the expression of average production
is as follow: the evolution matrix for initial distribution of the
demon is given by 
\[
\mathcal{T}=\mathcal{P}^{D}e^{\mathcal{R}\tau}\mathcal{M},\mathcal{\ P}^{D}=\begin{pmatrix}1 & 0 & 1 & 0\\
0 & 1 & 0 & 1
\end{pmatrix},\ \mathcal{M}=\begin{pmatrix}p_{0} & 0\\
0 & p_{0}\\
p_{1} & 0\\
0 & p_{1}
\end{pmatrix}
\]
Then the periodic steady state $\bm{p}_{\text{in}}^{D,ps}$ for demon
can be obtained by solving the linear equation 

\[
\mathcal{T}\bm{p}_{\text{in}}^{D,ps}=\bm{p}_{\text{in}}^{D,ps}.
\]
In the periodic steady state, the joint distribution of the demon
and the interacting bit, at the end of the interaction interval, is
given by $\bm{p}_{\tau}^{ps}=e^{\mathcal{R}\tau}\mathcal{M}\bm{p}_{\text{in}}^{D,ps}$.
The marginal distribution of the outgoing bit is then given by projecting
out the state of the demon:

\begin{equation}
\bm{p}_{\tau}^{B,ps}=(p_{0}^{f},p_{1}^{f})=\mathcal{P}^{B}e^{\mathcal{R}\tau}\mathcal{M}\bm{p}_{\text{in}}^{D},\mathcal{P}^{B}=\begin{pmatrix}1 & 1 & 0 & 0\\
0 & 0 & 1 & 1
\end{pmatrix}.
\end{equation}
Notice that $\bm{p}_{\text{in}}^{D,ps}$ is the first right eigenvector of
$\mathcal{T}$, $\bm{p}_{\tau}^{B,ps}=(p_{0}^{f},p_{1}^{f})$ then
can be solved, which determines the value of average production as
$\Phi=p_{1}^{f}-p_{1}$. By performing these calculations using Mathematica, 

\begin{equation}
\Phi=\frac{\delta-\epsilon}{2}\eta(\Lambda),\eta(\Lambda)=\frac{\nu_{2}P+\nu_{3}Q}{P+Q},
\end{equation}

\begin{equation}
P=\mu_{2}(\mu_{4}\nu_{3}+\mu_{1}\nu_{1}),Q=\mu_{3}(\mu_{4}\nu_{2}+\mu_{1}\nu_{1}),
\end{equation}
 where

\begin{equation}
\nu_{1}=1-e^{-2\Gamma\tau},\nu_{2}=1-e^{-(1+\Gamma-\alpha)\tau},\nu_{2}=1-e^{-(1+\Gamma+\alpha)\tau},
\end{equation}

\begin{equation}
\mu_{1}=(\delta+\sigma)\omega,\mu_{2}=\alpha+\Gamma+\sigma\omega,\mu_{3}=\alpha-\Gamma-\sigma\omega,\mu_{4}=1-\delta\omega
\end{equation}
 with $\alpha=\sqrt{1+\Gamma^{2}+2\Gamma\sigma\omega}$. 

\section{Spectral analysis of the matrix $\mathcal{T}$}

In this appendix, we gives some descriptions of the eigenvector expansion
method used in the main text,i.e., doing spectral analysis of the evolution matrix $\mathcal{T}$ for the demon. $\mathcal{T}$ has right eigenvectors $\bm{R}_{i}$

\begin{equation}
\mathcal{T}\bm{R}_{i}=\lambda_{i}\bm{R}_{i}
\end{equation}
and left eigenvectors $\bm{L}_{i}$ as 

\begin{equation}
\bm{L}_{i}^{\text{T}}\mathcal{\mathcal{T}}=\lambda_{i}\bm{L}_{i}^{\text{T}}
\end{equation}
with $\lambda_{i}$ the eigenvalues, which are sorted as $1=\lambda_{1}>\left|\lambda_{2}\right|\ge\left|\lambda_{3}\right|\ge...$
(we assume that $\lambda_{1}$ is not degenerate). The right eigenvector
$\bm{R}_{1}$ with $1=\lambda_{1}$ corresponds to the periodic steady
state, so we write $\bm{R}_{1}=\bm{p}_{\text{\text{in}}}^{D,ps}$.
According to the completeness relation, the initial state $\bm{p}_{\text{\text{in}}}^{D}$
can be expanded as 

\paragraph*{
\begin{equation}
\bm{p}_{\text{in,}0}=\bm{p}_{\text{in}}^{D,ps}+\sum_{i>1}d_{i}\bm{R}_{i},
\end{equation}
where
\begin{equation}
d_{i}=\frac{\bm{L}_{i}^{\text{T}}\cdot\bm{p}_{\text{in,0}}^{D}}{\bm{L}_{i}^{\text{T}}\cdot\bm{R}_{i}}.
\end{equation}
Calculation of the $i^{th}$ coefficient $d_{i}$}

For an arbitrary matrix $T$, it can be demonstrated that any pair
of left eigenvector and right eigenvector corresponding to different
eigenvalues of the matrix are mutually orthogonal. Here is the proof
(no degeneracy): 
\begin{align*}
T\bm{R}_{i} & =\lambda_{i}\bm{R}_{i}\Rightarrow\begin{cases}
\bm{L}_{j}^{\text{T}}T\bm{R}_{i} & =\lambda_{i}\bm{L}_{j}^{\text{T}}\bm{R}_{i}\\
\bm{L}_{j}^{\text{T}}T\bm{R}_{i} & =\lambda_{j}\bm{L}_{j}^{\text{T}}\bm{R}_{i}
\end{cases}\\
\Rightarrow & (\lambda_{i}-\lambda_{j})\bm{L}_{j}^{\text{T}}\bm{R}_{i}=0\\
\Rightarrow & \bm{L}_{j}^{\text{T}}\bm{R}_{i}=(\bm{L}_{i}^{\text{T}}\cdot\bm{R}_{i})\delta_{ij}.
\end{align*}
Therefore, for an evolution starting at a given initial distribution
$\bm{p}_{\text{in}}^{D}$, we have that $d_{i}$ is the corresponding overlap
coefficient between the initial probability and the $i^{th}$ left eigenvector
$\bm{L}_{i}^{\text{T}}$. During the relaxation process, the initial
distribution of the demon of the $n^{th}$ time interval, $\mathcal{T}^{n}\bm{p}_{\text{in,}0}^{D}$,
can be written as

\begin{equation}
\mathcal{T}^{n}\bm{p}_{\text{\text{in,0}}}^{D}=\bm{p}_{\text{in}}^{D,ps}+\sum_{i>1}d_{i}\lambda_{i}^{n}\bm{R}_{i}.
\end{equation}
 Then, 

\begin{equation}
\left\Vert \mathcal{T}^{n}\bm{p}_{\text{\text{in},0}}^{D}-\bm{p}_{\text{\text{in}}}^{D,ps}\right\Vert _{q}=\sum_{i>1}d_{i}\left\Vert \lambda_{i}\right\Vert _{q}^{n}\left\Vert R_{i}\right\Vert _{q}.
\end{equation}

The relaxation timescale is typically characterized by

\begin{equation}
\tau_{\text{rel}}=-\frac{1}{\ln\left|\lambda_{2}\right|}.
\end{equation}

It is then can be observed that a stronger effect (even shorter relaxation
time) can occur: a process where there exists a specific initial distribution
$\bm{\pi}_{\text{\text{in,0}}}^{D}$, such that 
\[
d_{2}|_{\bm{p}_{\text{\text{in},0}}^{D}=\bm{\pi}_{\text{\text{in,0}}}^{D}}=\frac{\bm{L}_{2}^{\text{T}}\cdot\bm{\pi}_{\text{in,0}}^{D}}{\bm{L}_{2}^{\text{T}}\cdot\bm{R}_{2}}=0.
\]

\section{Eigenvalues of the information refrigerator model}

Here we provide the expressions for the eigenvalues of the transition
matrices for the two-state demon model.

For the two-state demon, in the case of $\Gamma=1,\ \omega=1/2$ the
eigenvalues for $\mathcal{T}_{2\times2}$ reads

\begin{align}
\lambda_{1} & =1,\nonumber \\
\lambda_{2} & =\frac{e^{-2\tau}\left[-4+e^{s\tau}\left(-2+s \right)\left(-2+\delta \right)\left(-2+\epsilon \right)-e^{-s\tau}\left(2+\delta \right)\left(-2+\epsilon \right) + 4\delta\left(-2+\epsilon \right)+ 8\epsilon \right]}{4(4\epsilon-5)},\\
s & \equiv\sqrt{2+\frac{1-2\epsilon}{2-\epsilon}}
\end{align}
We draw the contour plot (which is not shown here) for $\lambda_{2}$ with $\tau=1$ and show
that $\lambda_{2}$ is positive for any value of $\epsilon\in[0,1/2]$
and $\delta\in[-1,1]$. It's obvious that the sign of $\lambda_{2}$
is irrelevant to the value of $\tau$, thus the necessary condition
for the existence of $N_{c}>0$, i.e. $\lambda_{2}>0$ can always be
satisfied in this model. 

\section{Derivation of the generalized second law of thermodynamics }

During any interaction interval, the joint distribution of the interacting
bit and the demon evolves according to the master equation
\[
\frac{d\boldsymbol{p}}{dt}=\mathcal{R}\boldsymbol{p}.
\]
Imagine that the interaction time $\tau$ is long enough ($\tau\rightarrow\infty$),
then the combined system will finally reach a steady state (the entries
of transition matrix $\mathcal{R}$ satisfy detailed balance condition,
so it's an equilibrium state)
\begin{equation}
\boldsymbol{p}^{ss}=\frac{(1,\mu,\mu\nu,\mu^{2}\nu)^{T}}{1+\mu+\mu\nu+\mu^{2}\nu},\mu=\frac{1+\sigma}{1-\sigma},\nu=\frac{1-\omega}{1+\omega},\label{eq:steady state}
\end{equation}
which makes $\mathcal{R}\boldsymbol{p}^{ss}=\boldsymbol{0}.$ When
$\tau\rightarrow\infty$, the steady state joint distribution $\boldsymbol{p}_{ss}$
is factorized as the product of marginal distributions $\boldsymbol{p}^{D,ss}$
and $\boldsymbol{p}^{B,ss}$, because the demon and bit are uncorrelated
at the beginning of each interval by construction. And in this case
distribution of bits can be regarded as an 'effective initial distribution'.
That is,
\begin{equation}
p_{ij}^{ss}=p_{i}^{D,ss}p_{j}^{B,ss},\ \ i\in\{u,d\},\ \ j\in\{0,1\},
\end{equation}
where $\boldsymbol{p^{D,ss}}=(1,\mu)^{T}/(1+\mu)$ and $\boldsymbol{p^{B,ss}}=(1,\mu\nu)^{T}/(1+\mu\nu).$
When interaction time $\tau$ is finite, the combined system always
relaxes towards this final steady state, though being interrupted
by the advent of new bits and stochastic resetting events (one should
note the similarity between resetting events for the demon and comings
of new bits). Therefore, the distribution of combined system will
get closer to the steady state $\boldsymbol{p}_{ss}$ at the end of
an interval, compared to its distribution at the start of the same
interval. That is, the relative entropy between any state $\boldsymbol{p}$
of the combined system and the steady state $\boldsymbol{p}_{ss}$
\begin{align}
D(\boldsymbol{p}||\boldsymbol{p}_{ss}) & =\sum_{k}p_{k}\ln\frac{p_{k}}{p_{k}^{ss}}\geq0,\label{eq:def_KL}\\
k\in & \{0u,0d,1u,1d\}
\end{align}
as a distance function is a Lyapunov function, satisfying 
\begin{equation}
\frac{d}{dt}D(\boldsymbol{p}||\boldsymbol{p}_{ss})\leq0\text{.}\label{eq:KL_divergence}
\end{equation}
Let $\boldsymbol{p}_{\text{in}}$ and $\boldsymbol{p}_{\tau}$ denote
the joint distribution at the start and at the end of a given interval
respectively, and similarly define $\boldsymbol{p}_{\text{in}}^{D}$,
$\boldsymbol{p}_{\tau}^{D}$, $\boldsymbol{p}_{\text{in}}^{B}$ and
$\boldsymbol{p}_{\tau}^{B}$ for the marginal distributions of the
demon and the bit. Then equation (\ref{eq:KL_divergence}) tells that
\begin{equation}
D(\boldsymbol{p}_{\text{in}}||\boldsymbol{p}_{ss})-D(\boldsymbol{p}_{\tau}||\boldsymbol{p}_{ss})\geq0,
\end{equation}
whose physical interpretation is the initial joint state is farther
from steady state than the final state at the end of the given interval
is. Note that the left hand side of the above equation is a standard
expression of conventional entropy production during a period of time
$\tau$ \cite{19PRL_informationbound}:
\[
\text{\ensuremath{\Sigma{}_{[0,\tau]}^{\textrm{tot}}\equiv}}D(\boldsymbol{p}_{\text{in}}||\boldsymbol{p}_{ss})-D(\boldsymbol{p}_{\tau}||\boldsymbol{p}_{ss})\geq0.
\]
Using (\ref{eq:steady state}) and (\ref{eq:def_KL}) one can rewrite
the above equation as
\begin{align}
S_{\tau}-S_{0}+\sum_{i\in\{u,d\}}\left(p_{\tau,i}^{D}-p_{\text{in},i}^{D}\right)\ln p_{i}^{D,ss}\nonumber \\
+\sum_{i\in\{0,1\}}\left(p_{\tau,i}^{B}-p_{\text{in},i}^{B}\right)\ln p_{i}^{B,ss} & \geq0,\label{eq:Shannon}
\end{align}
where $S_{0}$ and $S_{\tau}$ refer to the information entropies
of the joint distribution at the start and at the end of the current
interval. What we need to consider is just the new periodic steady
state in the presence of resetting. In this case, from the definition
of the average production with resetting $\phi_{r}$, $p_{\tau,1}^{B}-p_{\text{in},1}^{B}=-(p_{\tau,0}^{B}-p_{\text{in},0}^{B})=\phi_{r}$,
the last term of (\ref{eq:Shannon}) can be rewritten as 
\begin{align}
\sum_{i\in\{0,1\}}\left(p_{\tau,i}^{B}-p_{\text{in},i}^{B}\right)\ln p_{i}^{B,ss} & =\phi_{r}\ln\mu\nu\\
 & =Q_{c\rightarrow h}(\beta_{h}-\beta_{c})\text{.}
\end{align}
The joint information entropy $S$ can be decomposed as 
\begin{equation}
S=S_{D}+S_{B}+I(D;B),\ \ \ I(D;B)\geq0,
\end{equation}
where $S_{D}$ and $S_{B}$ are marginal information entropy of the
demon and the bit. Thus in our NPSS with resetting, equation (\ref{eq:Shannon})
gives (note that $I_{0}(D;B)=0$ due to the uncorrelated initial distribution)
\begin{equation}
Q_{c\rightarrow h}(\beta_{h}-\beta_{c})+\Delta S_{B}+\beta_{h}W_{\tau}^{r}\geq I_{\tau}(D;B)\geq0,
\end{equation}
where 
\begin{align}
\Delta S_{B}= & S_{B,\tau}-S_{B,0}\\
= & S_{B}(\delta')-S_{B}(\delta)=S_{B}(\delta-2\phi_{\text{tot}})-S_{B}(\delta),\nonumber \\
S_{B}(\delta)= & -\sum_{i=0}^{1}p_{i}\ln p_{i}\nonumber \\
= & -\frac{1-\delta}{2}\ln\frac{1-\delta}{2}-\frac{1+\delta}{2}\ln\frac{1+\delta}{2},\\
(\delta & =p_{0}-p_{1},\ \delta^{'}=p_{0}^{'}-p_{1}^{'})
\end{align}
and
\begin{align}
W_{\tau}^{r} & \equiv\Delta S_{D}+\sum_{i\in\{u,d\}}\left(p_{\tau,i}^{D,ps}-p_{\text{in},i}^{D,ps}\right)\ln p_{i}^{D,ss},\label{eq:r_work}\\
\Delta S_{D} & =-\sum_{i=d}^{u}p_{\tau,i}^{D,ps}\ln p_{\tau,i}^{D,ps}+\sum_{i=d}^{u}p_{\text{in},i}^{D,ps}\ln p_{\text{in},i}^{D,ps}
\end{align}
is the 'resetting work' during a whole interval $[n\tau,(n+1)\tau]$
due to stochastic resetting. In the original periodic steady state
without resetting, one has $\boldsymbol{p}_{\tau}^{D,ps}=\mathcal{T}\boldsymbol{p}_{\text{in}}^{D,ps}=\boldsymbol{p}_{\text{in}}^{D,ps}$
from definition of this state, so that the dissipated work (\ref{eq:r_work})
vanishes. However, in the NPSS $\boldsymbol{p}_{\tau}^{D,ps}(r)=\mathcal{T}\boldsymbol{p}_{\text{in}}^{D,ps}(r)\neq\boldsymbol{p}_{\text{in}}^{D,ps}(r)$
according to the definition (\ref{eq:new_pss}). In the NPSS, one
can obtain
\begin{align}
\boldsymbol{p}_{\tau}^{D,ps}(r)-\boldsymbol{p}_{\text{in}}^{D,ps}(r)=\mathcal{T}\boldsymbol{p}_{\text{in}}^{D,ps}(r)-\boldsymbol{p}_{\text{in}}^{D,ps}(r)\\
=\frac{1-e^{-r\tau}}{1-\lambda_{2}e^{-r\tau}}d_{2}\left(\mathcal{T}\bm{R}_{2}-\bm{R}_{2}\right) & .
\end{align}
Because $p_{\tau,u}^{D,ps}-p_{\text{in},u}^{D,ps}=-(p_{\tau,d}^{D,ps}-p_{\text{in},d}^{D,ps})\equiv\Delta p^{D,ps}$,
the second contribution of resetting work can be written as 
\begin{align}
\sum_{i=d}^{u}\left(p_{\tau,i}^{D,ps}-p_{0,i}^{D,ps}\right)\ln p_{i}^{D,ss} & =\Delta p^{D,ps}\ln\mu\nonumber \\
 & =\beta_{h}\Delta p^{D,ps}\Delta E,
\end{align}
thus the resetting work during each interval in the NPSS is given
by 
\begin{equation}
W_{\tau}^{r}=\Delta S_{D}+\beta_{h}\Delta p^{D,ps}\Delta E.
\end{equation}
The above modified second law can also be derived from an integral
fluctuation theorem for the stochastic entropy production. Following
Seifert's spirit, the total stochastic entropy production in an interaction
interval for a single trajectory $\Gamma_{i\rightarrow j}$ starting
in state $i$ at initial time and ending in state $j$ at time $\tau$
is defined as ($i,j\in\{0u,0d,1u,1d\}$)
\begin{align}
\sigma(\Gamma_{i\rightarrow j})= & \ln\frac{p_{i}(0)p(\Gamma_{i\rightarrow j})}{p_{j}(\tau)p(\Gamma_{j\rightarrow i}^{\dagger})},\\
= & \ln\left[\frac{p_{i}(0)}{p_{j}(\tau)}\prod_{k,l\in\Gamma_{i\rightarrow j}}\left(\frac{R_{kl}}{R_{lk}}\right)^{n_{kl}}\right],
\end{align}
which naturally gives rise to an integral fluctuation theorem
\begin{align}
\langle e^{-\sigma(\Gamma_{i\rightarrow j})}\rangle= & \sum_{i,j}\int d\Gamma_{i\rightarrow j}p_{i}(0)p(\Gamma_{i\rightarrow j})e^{-\sigma(\Gamma_{i\rightarrow j})}\nonumber \\
= & \sum_{i,j}\int d\Gamma_{i\rightarrow j}p_{j}(\tau)p(\Gamma_{j\rightarrow i}^{\dagger})\nonumber \\
= & \sum_{i,j}\int d\Gamma_{j\rightarrow i}^{\dagger}p_{j}(\tau)p(\Gamma_{j\rightarrow i}^{\dagger})=1.
\end{align}
Then using Jensen equality, it follows that
\[
\Sigma{}_{[0,\tau]}^{\textrm{tot}}\equiv\langle\sigma(\Gamma_{i\rightarrow j})\rangle\geq0.
\]
It has been proven that \cite{Seifert2012Stochastic}
\begin{equation}
\langle\frac{d}{dt}\sigma(\Gamma_{i\rightarrow j})\rangle=\left[R_{ij}p_{j}(t)-R_{ji}p_{i}(t)\right]\ln\frac{R_{ij}p_{j}(t)}{R_{ji}p_{i}(t)}
\end{equation}
Then we have 
\begin{align*}
\langle\sigma(\Gamma_{i\rightarrow j})\rangle= & \int_{0}^{\tau}\sum_{i>j}\left[R_{ij}p_{j}(t)-R_{ji}p_{i}(t)\right]\ln\frac{R_{ij}p_{j}(t)}{R_{ji}p_{i}(t)}\\
= & -\int_{0}^{\tau}\frac{d}{dt}D(\boldsymbol{p}(t)||\boldsymbol{p}_{ss})\\
= & Q_{c\rightarrow h}(\beta_{h}-\beta_{c})+\Delta S_{B}+\beta_{h}W_{\tau}^{r}-I_{\tau}(D;B)\geq0,
\end{align*}
where in the second line the detailed balance condition $R_{ij}p_{j}^{ss}=R_{ji}p_{i}^{ss}$
has been used \cite{2017England_Entropy}.

\end{document}